\documentstyle[preprint,aps,epsfig]{revtex}

\def\tb{\tilde{b}}
\begin{document}
\draft

\title{Nuclear Effects on Heavy Boson Production at RHIC and LHC}

\author{Xiaofei Zhang\thanks{electronic mail: xiaofei@cnr4.physics.kent.edu}
 and George Fai\thanks{electronic mail: fai@cnr4.physics.kent.edu}}

\address{Center for Nuclear Research, Department of Physics,
Kent State University \\
Kent, Ohio 44242, USA}

\maketitle

\vspace*{-6.5cm}
\begin{flushright}
{KSUCNR-201-02}
\end{flushright}
\vspace*{5.5cm}

\begin{center}
\today
\end{center}

\begin{abstract}

We predict $W^\pm$ and $Z^0$ transverse momentum distributions from
proton-proton  and nuclear collisions at RHIC and LHC.
A resummation formalism with power corrections to the renormalization 
group equations is used. The dependence of the resummed 
QCD results on the non-perturbative input is very weak
for the systems considered. Shadowing effects  are discussed and found 
to be unimportant at RHIC, but important for LHC. We study
the enhancement of power corrections due to multiple
scattering in nuclear collisions and numerically illustrate
the weak effects of the dependence on the nuclear 
mass.

\end{abstract}
\vspace{0.2in}


\section{Introduction}
\label{sec1}

With the commissioning of the Relativistic Heavy Ion Collider (RHIC),
nuclear collision physics entered the collider era.
New phenomena have been presented in $Au + Au$ collisions
at the center-of-mass energy $\sqrt{s}=$ 130 GeV/nucleon\cite{QM01}. 
Negative hadron multiplicities are significantly increased compared 
to proton-antiproton collisions\cite{star}.
Some of the  data, such as the  antiproton-to-proton 
ratios\cite{phenix1},
are reported to contradict perturbative Quantum Chromodynamics (pQCD)
calculations. Initial efforts to understand hadron spectra from
nuclear reactions at RHIC using pQCD have been published
recently\cite{wang01,zfpbl02}. However, there is still a strong
debate on how to apply pQCD to hadron production
even in proton-proton ($pp$) collisions.
With RHIC running at the full energy, and experiments at the Large
Hadron Collider (LHC) at several TeV/nucleon soon to follow,
``clean'' processes, where pQCD works well in $pp$ collisions,
are needed to test pQCD in nuclear reactions at collider energies.
It is particularly important to distinguish between
expected nuclear effects and ``new physics''.

The production of $W^\pm$ and $Z^0$ bosons has been extensively studied
at Tevatron energy ($\sqrt s=1.8$ TeV). Perturbative QCD proved to be 
very successful in
explaining the CDF\cite{CDF-Z} and D0\cite{D0-W} data. Due to the large mass
of $W^\pm$ and $Z^0$, and no final state rescattering in their production 
process, power corrections are expected to be small. Therefore,
$W^\pm$ and $Z^0$ production could provide a bench mark test for pQCD
at RHIC and LHC in both $pp$ and nuclear collisions.  Luminosity
remains a concern at RHIC, in particular in $pp$ collisions.

The bulk of the data for $W^\pm$ and $Z^0$ production is concentrated
in the small transverse momentum ($p_T$) region, where $p_T$ is much
smaller than the corresponding heavy boson mass ($M$).
When $p_T \ll M$, the $p_T$ distributions calculated order-by-order in
$\alpha_s$ in conventional fixed-order perturbation theory
receive a large logarithm, $\ln(M^2/p_T^2)$, at every power of
$\alpha_s$, even in the leading order in $\alpha_s$.
Therefore, at sufficiently small $p_T$, the convergence of the
conventional perturbative expansion in powers of $\alpha_s$ is
impaired, and the logarithms must be resummed.

Resummation of the large logarithms in QCD can be carried out either
in $p_T$-space directly, or in the so-called
``impact parameter'', $\tb$-space, which
is a Fourier conjugate of the $p_T$-space. (Since we are also interested
in {\it nuclear} collisions, we reserve the notation $b$ for the usual
impact parameter associated with the geometry of colliding heavy nuclei,
and denote the mathematical tool used in the resummation procedure
by the symbol $\tb$.) Using the renormalization group
equation technique, Collins and Soper improved the $\tb$-space
resummation to resum all logarithms as singular as,
$\ln^m(M^2/p_T^2)/p_T^2$, as $p_T\rightarrow 0$ \cite{CS-b}. In the
frame work of this renormalization group improved $\tb$-space
resummation, Collins, Soper, and Sterman (CSS) derived a formalism
for the transverse momentum distribution of vector boson production
in hadronic collisions\cite{CSS-W}. In the CSS formalism, non-perturbative
input is needed for the large $\tb$ region.  The dependence of the pQCD
results on the non-perturbative input is not weak if the original
extrapolation proposed by CSS is used. Recently, a new extrapolation
scheme was proposed, based on solving the renormalization group
equations, including power corrections\cite{qz01}. Using
the new extrapolation formula, the dependence of the pQCD result
on the non-perturbative input was significantly reduced.
This was achieved without matching the fixed-order calculations. The
results agree with the D0 nd CDF data very well in the entire $p_T$
interval from $p_T \lesssim 1$ GeV to $p_T$ as large as the
the vector mass. In addition, in this new extrapolation
formalism, power corrections can be studied systematically for $pp$ to
proton-nucleus ($pA$) to nucleus-nucleus ($AB$) collisions.

Factorization plays a key role in pQCD predictions of observables.
At RHIC and LHC, nuclear collisions reach an unprecedented energy scale.
Factorization can not be extrapolated easily to $AB$ collisions
from hadron-hadron collisions\cite{qiu-sterman-fac},
and power corrections may become important \cite{lq2}.
There are two types of power corrections: (1) Power corrections directly
to the physical observables (type-I) and (2) power corrections to the
evolution of renormalization group equations (type-II). Type-I power
correction are proportional to powers of $(\Lambda_{QCD}/Q)$, Q being the
physical large scale (momentum). These are small for $W^\pm$ and $Z^0$
production as a result of the large mass of the particles in question.
Type-II power corrections are proportional
to powers of $(\Lambda_{QCD}/\mu)$, with evolution scale $\mu$.
Therefore, physical observables carry the effect of type-II power corrections
for all $\mu[Q_0,Q]$, with the boundary condition at the scale $Q_0$.
Even with large mass, $W^\pm$ and $Z^0$ can still carry a large effect 
of type-II power corrections. However, Ref. \cite{qz01} provides a
way to study the effect of type-II power corrections numerically.
Based on the numerical results presented in Ref. \cite{qz01},
the effect of type-II power corrections is also expected to be 
small for heavy vector boson production
in $pp$ collisions. In nuclear collisions, type-II power corrections
will be enhanced by the nuclear size. Here we study the effects
of enhanced power correction in $pA$ and $AB$ collisions.

The rapidity distribution of $W^\pm$ and $Z^0$ bosons at LHC was calculated
recently and the effect of shadowing was discussed in this
context\cite{vogt01}. The decay spectrum of $Z^0$ was suggested as an
alternative reference process for $J/\psi$ suppression at LHC,
since the Drell-Yan lepton pairs at low mass are expected to
be dominated by $b\bar{b}$ decay. As most of the $W^\pm$ and $Z^0$ 
bosons observed in the experiment will be in the small transverse
momentum region, it is of great interest to study
the nuclear effects on massive gauge boson transverse momentum
distributions. These effects include both the nuclear modification of the 
parton distribution function
(shadowing), and the initial state power corrections.
The second part of the present paper is devoted to these subjects.

The rest of the paper is organized as follows. In Section \ref{sec2},
we review the CSS $\tb$-space resummation formalism and the new
procedure\cite{qz01} to extrapolate the pQCD calculation to the large $\tb$
region is outlined. We give predictions for $W^\pm$ and $Z^0$ transverse 
momentum distributions from $pp$ collision at both RHIC and LHC energies.
The sensitivity of the prediction to the non-perturbative parameters
is discussed. Nuclear effects on $W^\pm$ and $Z^0$ transverse momentum
distributions in $pA$ and $AB$ collisions are studied in Sections
\ref{sec3} and \ref{sec4}. In Section \ref{sec5}, we present our
conclusions. To make the paper reasonably self-contained, the
Appendix summarizes relevant details of the CSS formalism.
We use $\hbar = c = 1$ units.



\section {Transverse momentum distribution of heavy vector bosons
in proton-proton collisions}
\label{sec2}

In this Section, we briefly review the CSS $\tb$-space resummation
formalism (Section \ref{subsec21}). In Section \ref{subsec22}
we introduce the recently-proposed new formalism to extrapolate
the perturbative calculation to the large $\tb$ region\cite{qz01},
and discuss the transverse momentum distribution of heavy vector bosons
in $pp$ collisions at RHIC and LHC energies.

\subsection{ Collins-Soper-Sterman (CSS) formalism}
\label{subsec21}

In the production of heavy vector bosons (of mass $M$),
when $p_T \ll M$, the $p_T$ distribution calculated order-by-order in
$\alpha_s$ in conventional fixed-order perturbation theory
receives a large logarithm, $\ln(M^2/p_T^2)$, at every power of
$\alpha_s$. Even at the leading order in $\alpha_s$, the cross
section $d\sigma/dM^2\,dy\,dp_T^2$ contains a term proportional to
$(\alpha_s/p_T^2)\ln(M^2/p_T^2)$ coming from the partonic subprocess
$q+\bar{q} \rightarrow V +g$, where $V=\gamma^*,W^{\pm}$ or $Z^0$.  Beyond 
the leading order, we actually get two powers of the logarithm for
every power of $\alpha_s$, due to soft and collinear gluons emitted by
the incoming partons.
Therefore, at sufficiently small $p_T$, the convergence of the
conventional perturbative expansion in powers of $\alpha_s$ is
impaired, and the logarithms must be resummed.

For vector boson production in hadronic collisions
between hadrons $h_A$ and $h_B$, $h_A + h_B \rightarrow V(M)+X$ with
$V=\gamma^*,W^{\pm}$, or $Z^0$, the CSS resummation formalism yields the
following generic form \cite{CSS-W}:
\begin{equation}
\frac{d\sigma(h_A+h_B\rightarrow V+X)}{dM^2\, dy\, dp_T^2} =
\frac{1}{(2\pi)^2}\int d^2 \tb\, e^{i\vec{p}_T\cdot \vec{\tb}}\,
\tilde{W}(\tb,M,x_A,x_B) + Y(p_T,M,x_A,x_B) \,\,\, ,
\label{css-gen}
\end{equation}
where $x_A= e^y\, M/\sqrt{s}$ and $x_B= e^{-y}\, M/\sqrt{s}$, with
rapidity $y$ and collision energy $\sqrt{s}$.
In Eq.~(\ref{css-gen}), the $\tilde{W}$ term dominates the $p_T$ distributions
when $p_T \ll M$, and the $Y$ term gives corrections that are negligible
for small $p_T$, but become important when $p_T\sim M$. In the CSS formalism,
the function $\tilde{W}$ appears as a superposition,
\begin{equation}
\tilde{W}(\tb,M,x_A,x_B) = \sum_{ij}
\tilde{W}_{ij}(\tb,M,x_A,x_B)\, \sigma_{ij\rightarrow V}(M) \,\,\, ,
\label{css-W-ij}
\end{equation}
where $\sigma_{ij\rightarrow V}(M)$ is the lowest order cross section
for a quark-antiquark pair of invariant mass $M$ to annihilate
into a vector boson $V$, and the $\sum_{ij}$ runs over all
possible quark and antiquark flavors that can annihilate into the
vector boson at the Born level\cite{CSS-W}. In Eq.~(\ref{css-W-ij}),
$\tilde{W}_{ij}(\tb,M,x_A,x_B)$ is an effective flux to have partons of
flavor $i$ and $j$ from the respective hadrons $h_A$ and $h_B$, and it
has the form
\begin{equation}
\tilde{W}_{ij}(\tb,M,x_A,x_B) =
{\rm e}^{S(\tb,M)}\, \tilde{w}_{ij}(\tb,c/\tb,x_A,x_B) \,\,\, ,
\label{css-W-sol}
\end{equation}
where $S(\tb,M)$ is given in the Appendix, and $c$ is a constant of order
unity\cite{CSS-W,qz01}. The functions $\tilde{w}_{ij}(\tb,c/\tb,x_A,x_B)$ in
Eq.~(\ref{css-W-sol}) depend on only one momentum scale, $1/\tb$, and
are perturbatively calculable as long as $1/\tb$ is large
enough. All large logarithms from $\ln(1/\tb^2)$ to $\ln(M^2)$ in
$\tilde{W}_{ij}(\tb,M,x_A,x_B)$ are completely resummed into the
exponential factor $\exp[S(\tb,M)]$.

Since the perturbatively resummed $\tilde{W}_{ij}(\tb,M,x_A,x_B)$ in
Eq.~(\ref{css-W-sol}) is only reliable for the small $\tb$ region, an
extrapolation to the large $\tb$ region is necessary in order to
complete the Fourier transform in Eq.~(\ref{css-gen}).
This is the point where the dependence of the result on the nonperturbative
parameters enters into the formalism, possibly limiting the predicative power
of pQCD , especially for small $p_T$. We will next discuss this in
detail.

\subsection{Improved extrapolation formula with power corrections}
\label{subsec22}

The CSS formalism in its original form introduces a
modification to the output of the perturbative calculation (see
Appendix, Eq. (\ref{css-W-b})).
The size of the modification depends on the non-perturbative input 
parameters, and can be as large as 50\% of the perturbative result.
Therefore the predictive power of the CSS formalism was questioned 
and new efforts have been devoted to resum the large logarithms directly
in $p_T$ space \cite{Ellis-1,Ellis-2}.

In oder to improve the predictive power of the CSS $\tb$-space
resummation formalism, the following new form was proposed\cite{qz01}
by solving the renormalization group equation including power
corrections:
\begin{equation}
\tilde{W}(\tb,M,x_A,x_B) = \left\{
\begin{array}{ll}
\tilde{W}(\tb,M,x_A,x_B) & \quad \mbox{$\tb\leq \tb_{max}$} \\
\tilde{W}(\tb_{max},M,x_A,x_B)\,
F^{NP}(\tb,M,x_A,x_B;\tb_{max},\alpha)
& \quad \mbox{$\tb > \tb_{max}$}
\end{array} \right. \,\, ,
\label{qz-W-sol-m}
\end{equation}
where the
nonperturbative function $F^{NP}$ in Eq.~(\ref{qz-W-sol-m}) is
given by
\begin{eqnarray}
F^{NP}(\tb,M,x_A,x_B;\tb_{max},\alpha)
& = &
\exp\Bigg\{ -\ln\left(\frac{M^2 \tb_{max}^2}{c^2}\right) \left[
g_1 \left( (\tb^2)^\alpha - (\tb_{max}^2)^\alpha\right)
+g_2 \left(\tb^2 - \tb_{max}^2\right) \right]
\nonumber \\
&\ & {\hskip 0.8in}
-\bar{g}_2 \left(\tb^2 - \tb_{max}^2\right) \Bigg\} \,\,\, .
\label{qz-fnp-m}
\end{eqnarray}

In Eq. (\ref{qz-W-sol-m}) $\tb_{max}$ is a parameter to separate 
the perturbatively calculated part from the non-perturbative
input. Unlike in the original CSS formalism,
when $\tb \leq \tb_{max}$, the perturbatively calculated
$\tilde{W}(\tb,M,x_A,x_B)$ is not altered and is independent of 
the nonperturbative parameters. In addition, the $\tb$-dependence in 
Eq.~(\ref{qz-fnp-m}) is separated according to different physics
origins. The $(\tb^2)^\alpha$-dependence mimics the
summation of the perturbatively calculable leading power
contributions to the kernels $K$ and $G$ in Eq.s (\ref{css-K-rg}) and
(\ref{css-G-rg}) of the Appendix, respectively, to all orders in the running
coupling constant $\alpha_s(\mu)$ with the scale $\mu$ running into
the nonperturbative region \cite{AMS-DY}. The $\tb^2$-dependence of the
$g_2$ term is a direct consequence of dynamical power corrections to the
renormalization group equations of the kernels $K$ and $G$, and has an 
explicit dependence on $M$. The $\bar{g}_2$ term represents the effect 
of the non-vanishing intrinsic parton transverse momentum.
As we discuss later, these two terms behave differently in nuclear
collisions.

A remarkable feature of the $\tb$-space resummation formalism is
that the resummed exponential factor $\exp[S(\tb,M)]$
(see Eq.~(\ref{css-W-sol})) suppresses the
$\tb$-integral when $\tb$ is larger than $1/M$. Therefore, when $M\gg \mu_0$,
 (where $\mu_0 \sim 1/\tb_{max}$ represents the scale at which pQCD starts to 
be applicable), it is possible that the Fourier transform in
Eq.~(\ref{css-gen}) is dominated by a region of $\tb$ much smaller than
$1/\mu_0$, and the calculated $p_T$ distributions are insensitive to
the nonperturbative information at large $\tb$. In fact, it was shown
using the saddle point method that, for a large
enough $M$, a QCD perturbation calculation is valid even 
at $p_T=0$\cite{PP-b,CSS-W}.
As discussed in Ref. \cite{qz01}, the value of the saddle point
strongly depends on the
collision energy $\sqrt{s}$, in addition to its well-known $M^2$
dependence. Because of the steep evolution of
parton distributions at small $x$, the $\sqrt{s}$ dependence of the
$\tilde{W}$ in Eq.~(\ref{css-gen}) significantly decreases the value of
the saddle point and improves the
predictive power of the $\tb$-space resummation formalism at collider
energies.

To display the saddle point clearly, let us rewrite the first term 
on the right-hand side of  Eq.~(\ref{css-gen}),
taking into account that there is no preferred transverse direction 
and that $\tilde{W}$ in Eq.~(\ref{css-gen})
is a function of $\tb=|\vec{\tb}|$ only: 
\begin{equation}
\frac{1}{(2\pi)^2}\int d^2\tb\, e^{i\vec{p}_T\cdot \vec{\tb}}\,
\tilde{W}(\tb,M,x_A,x_B) =
\frac{1}{2\pi} \int_0^\infty d\tb\, \tb\, J_0(p_T \tb)\,
\tilde{W}(\tb,M,x_A,x_B)  \,\,  ,
\label{css-W-F}
\end{equation}
where $\tilde{W}$ contains the exponential factor $\exp[S(\tb,M)]$, and
$J_0(z)$ with $z=p_T \tb$ represents the Bessel function of order zero.
When $p_T > 0$, the Bessel function suppresses
the large $\tb$ region of the integration. Because the argument of
the Bessel function is proportional to $p_T$, the large $\tb$ region is
more suppressed if $p_T$ is larger. In the following, we focus on
the saddle point at $p_T=0$. Fig.~1 is a graph (on an arbitrary scale)
of the integrand of the right hand side of Eq.~(\ref{css-W-F})
with $p_T=0$ for the case of $Z^0$ production at the collision energies 
where we intend to carry out our calculations: 
$\sqrt{s} =$ 350~GeV (dotted line), $\sqrt{s} =$ 500 GeV (short dashes), 
and at $\sqrt{s} =$ 5.5 TeV (solid). The position of the saddle point
decreases as the collision energy increases. We will show that the 
predictive power of the formalism is very good at RHIC.
As a consequence of the behavior of the saddle point position,
we expect excellent precision at $\sqrt{s} =$ 5.5 TeV.
At the LHC $pp$ energy of $\sqrt{s} =$ 14 TeV,
the saddle point moves to even smaller values of $\tb$, improving
the accuracy of the calculation even further.

The parameters $g_1$ and $\alpha$ of Eq.~(\ref{qz-fnp-m})
are fixed by the requirement of continuity of the function
$\tilde{W}(\tb)$  and its derivative at $\tb=\tb_{max}$.
(The results are insensitive to changes of $\tb_{max}$ in
the interval 0.3 GeV$^{-1} \lesssim \tb_{max} \lesssim$ 0.7 GeV$^{-1}$.
We use $\tb_{max}=$ 0.5 GeV$^{-1}$ in Ref. \cite{qz01} and here.) 
The value of $g_2$ and ${\bar g}_2$ can be obtained by fitting the
low-energy Drell-Yan data, and are taken to be
${\bar g}_2\sim 0.27$ GeV$^2$, and $g_2 \sim 0.01$ GeV$^2$ following 
Ref.s \cite{qz01} and 
\cite{zhang-fai}.
As the $\tb$ dependence of the  $g_2$ and ${\bar g}_2$ terms in
Eq.~(\ref{qz-fnp-m}) is identical, it is convenient to combine
these terms and define
\begin{equation}
G_2= \ln({M^2 \tb_{max}^2\over {c^2}})g_2 + \bar{g}_2 \,\,\, .
\label{G2}
\end{equation}
Using the values of the parameters listed above, we get $G_2 \sim 0.4$ GeV$^2$
 for $W^\pm$ and $Z^0$ production in $pp$ collisions. 
The parameter $G_2$ can be considered 
the only free parameter in the non-perturbative input in Eq. (\ref{qz-fnp-m}),
arising from the power corrections in the renormalization group equation.
Our results are not sensitive to small variations of $G_2$ around
its estimated value. We use $G_2=0.4$ GeV$^2$ in the following.
An impression about the importance of power corrections can be obtained by
comparing results with $G_2=0.4$ GeV$^2$ 
to those with power corrections turned off ($G_2=0$).
We therefore define the ratio of these quantities:
\begin{equation}
R_{G_2}(p_T) \equiv \left.
\frac{d\sigma^{(G_2)}(p_T)}{dp_T} \right/
\frac{d\sigma(p_T)}{dp_T} \,\,\, .
\label{Sigma-g2}
\end{equation}

The cross sections in the above equation and in the results presented
in this paper have been integrated over rapidity and invariant mass squared
from Eq. (\ref{css-gen}). For the integration over $dM^2$, we use the
``narrow width approximation''\cite{pinkbook}. With respect to rapidity, 
we integrate over the interval $[-1, 1]$ for RHIC, and over [-2.4,2.4] 
for LHC, representing the central acceptances of the appropriate detectors.
For the parton distribution functions, we use the CTEQ5M set\cite{CTEQ5} 
in the present work.

The following figures display differential cross sections 
and the corresponding $R_{G_2}$ ratios for $Z^0$ and $W^\pm$ 
production as functions of $p_T$, in order of increasing
energy. (In what follows, by $W^\pm$ production cross section
we mean the cross section for $W^+ + W^-$.)
In Fig.~2(a) we show the differential cross section for  
$Z^0$ production in $pp$ collisions at $\sqrt s= 500$ GeV,   
where a RHIC $pp$ run is expected. The accompanying 
Fig.~2(b) displays $R_{G_2}$ of Eq.~(\ref{Sigma-g2}). It 
can be seen that the $R_{G_2}$ ratio converges to unity 
(i.e. very small power corrections) with increasing $p_T$, and 
the effect of power corrections is about five percent 
for $p_T=0$. However, when $p_T>2$~GeV, the deviation of
$R_{G_2}$ from unity is under two percent. A very similar 
conclusion can be drawn from Fig.~3, concerning $W^\pm$ production 
in $pp$ collisions at the same energy: the dependence on the 
nonperturbative input is weak even at the lowest transverse momenta.

Fig.-s 4 and 5 contain similar information for $Z^0$ and $W^\pm$ production 
in $pp$ collisions at the planned LHC nucleus-nucleus
collision energy of $\sqrt s= 5.5$ TeV,
while the cross section and $R_{G_2}$ are given
for the LHC $pp$ energy of $\sqrt s=14$ TeV in Fig.-s~6 and 7.
At all energies, the magnitude of the $W^\pm$ production cross section 
is about 3 to 5 times that of the $Z^0$ production cross section.
Otherwise, the results for $Z^0$ and $W^\pm$ production are very 
similar, because the saddle points are almost at the same 
place in $\tb$ space in the two cases. As it has been pointed out 
in Ref\cite{qz01}, the position
of the saddle point is determined by two terms: the first  term 
is inversely proportional to the mass of the vector boson,  
and the second term is proportional to the derivative 
of the parton distribution function with respect to 
the scale at $x_A$ or  $x_B$.
At the collider energies considered here, the second term is negative.  
While $M_W$  is  a little smaller than $M_Z$, and thus the first term
is larger for $W^{\pm}$ than for $Z^0$, the contribution from the 
second term will effect the saddle point for $W^\pm$ more 
than that of the $Z^0$ with a small $x_A$ and $x_B$.
   
As expected from Fig.~1, the dependence on the nonperturbative input 
decreases with increasing collision energy. The $R_{G_2}$ ratio
is smaller than one percent at LHC for both $\sqrt{s}=5.5$~TeV
and $\sqrt{s}=14$ TeV, even when $p_T=0$. These results  confirm
that the predictive power of the formalism is  very good at RHIC
and excellent at LHC.  They also show that the effect of power corrections
is very small at LHC for almost the whole $p_T$  region and it is also
negligible for RHIC when $p_T$ is larger than $2$ GeV.



\section{Isospin and shadowing effects in nuclear collisions}
\label{sec3}

In lack of nuclear effects on the hard collision,
the production of heavy vector bosons in nucleus-nucleus ($AB$) collisions
should scale, compared to the production in $pp$ collisions,
as the number of hard collisions, $AB$. 
However, there are several nuclear effects on the hard collision in a
heavy-ion reaction. We distinguish three categories of these effects:
isospin effects, the modification of the parton distribution
function in the nucleus, and the enhancement of power corrections 
by the nuclear size. In this Section we discuss the first two 
of these nuclear effects. Power corrections will be treated 
separately in the next Section.

As the parton distribution of neutrons is different
from that of the protons, the production cross section
of heavy vector bosons in proton-neutron and neutron-neutron interactions 
differs from the corresponding production cross section
in $pp$ collisions. This difference is the source of the
so-called isospin effects. To study the isospin effects, 
let us introduce
\begin{equation}
R_{iso}(p_T) \equiv \left.
\frac{d\sigma(p_T,Z_A/A,Z_B/B)}{dp_T} \right/ 
\frac{d\sigma(p_T)}{dp_T} \,\,\,  ,
\label{sigma_iso}
\end{equation}
where $Z_A$ and $Z_B$ are the atomic numbers and $A$ and $B$ are 
the mass numbers of the colliding nuclei, and the cross section
$d\sigma(p_T,Z_A/A,Z_B/B)/dp_T$ has been averaged over $AB$,
while $d\sigma(p_T)/dp_T$ is the $pp$ cross section (with $G_2 = 0$). 
Since the $u \bar u $ coupling in $Z^0$ production is replaced by 
$u \bar d $ coupling to produce $W^\pm$ bosons, we expect $R_{iso}$
to be complementary for $W^\pm$ and $Z^0$ production, i.e. if one of 
these ratios is larger than one, the other ratio should be smaller than 
unity.

Next we turn to the phenomenon of shadowing. We use the term
shadowing in the general sense, referring to all
modifications of the parton distribution function 
in the environment of the nucleus\cite{vogt01}. 
This includes, in different regions of
$x$, the phenomena of shadowing (in the strict sense),
anti-shadowing, and the EMC effect. It is also important to note that
we define shadowing effects as leading twist effects, 
not including power corrections\cite{qiu-sterman-fac}. Thus, 
shadowing effects and power correction are clearly separated
in the present work.

In general, shadowing is expected to be a function of $x$,
the scale $Q$, and of the position in the nucleus. The latter
dependence means that in heavy-ion collisions, shadowing
should be impact parameter ($b$) dependent. The
parameterizations of shadowing in the literature take 
into account some of these effects, but no complete parameterization 
exists to date to our knowledge. For example, the HIJING 
parameterization includes impact parameter
dependence, but does not deal with the scale dependence\cite{wang1,wang2}.
On the other hand, the EKS98\cite{eks} and HKM\cite{HKM}
parameterizations have a scale dependence,
but do not consider impact parameter dependence.
(The latter parameterizations have been compared recently\cite{eskola02}.)
Since in this paper we concentrate on impact-parameter integrated 
results, where the effect of the $b$-dependence of shadowing is relatively 
unimportant\cite{zfpbl02}, and we focus more attention on scale dependence, 
we find the EKS98 shadowing parameterization most appropriate for our present 
purposes. Therefore we use EKS98 shadowing\cite{eks} in this work.

Similarly to Eq. (\ref{sigma_iso}) for isospin effects, 
we define a shadowing ratio,
\begin{equation}
R_{sh}(p_T) \equiv \left.
\frac{d\sigma^{(sh)}(p_T,Z_A/A,Z_B/B)}{dp_T} \right/
\frac{d\sigma(p_T)}{dp_T} \,\,\, .
\label{Sigma-sh}
\end{equation}
Thus, $R_{sh}$ incorporates the effects of both shadowing and the 
isospin composition of the nuclei. Furthermore, in this Section we 
take $G_2=0$ throughout, switching off power corrections.
The enhanced power corrections in $pA$
and $AB$ collisions will be discussed separately in 
the next Section.

Fig. 8 displays $R_{iso}$ (dashed lines) and $R_{sh}$ (full lines)
as functions of $p_T$ for $Z^0$ production (Fig. 8(a)) and $W^\pm$ 
production (Fig. 8(b)), respectively, at $\sqrt{s}=$ 350 MeV. 
The  same information is presented at $\sqrt{s}=$ 5.5 TeV in Fig. 9. 
We find that $R_{iso} > 1$  for $Z^0$ production and $R_{iso} < 1$  
for $W^\pm$ production at both energies, complementing each other, as  
expected. How close  $R_{iso}$ is to unity
depends on the region of $x$ that has the dominant contribution to 
the cross section. At LHC,  $x \sim 0.02$, and
the magnitude of the isospin effects is about 2\%. This is because
when $x$ is in this range, the difference of the distributions of 
$u(\bar u)$ and $d(\bar d)$ quarks in the nucleus is very small.
At RHIC, where the dominant $x \sim 0.26$, the isospin effects
are about 20\% in the small $p_T$  region. We have also 
checked that the scale dependence of isospin effects is not important. 
In Fig.~8, we also show a dotted line, which is the ratio of 
$R_{sh}$ to $R_{iso}$, and thus can be considered  to display
the effect of ``pure shadowing''. Pure shadowing appears as an 
approximately constant reduction of 3-5\% at  $x \sim 0.26$.

The appearance of $R_{sh}$ (full lines) is more surprising 
at $\sqrt{s}=$ 5.5 TeV (Fig.~9). Since at LHC,  
even at $p_T=90$ GeV, $x\sim 0.05$, we are still in the ``strict
shadowing'' region. Therefore, the fact that 
$R_{sh} > 1$  for 20 GeV $\lesssim p_T \lesssim$ 70 GeV is 
not ``antishadowing''; rather, it is a consequence of the
change of the shape of the cross section from $pp$ to $AB$
reactions. In $pp$ collisions, the maximum of the cross section 
is at $p_T \approx$ 3.7 GeV. In $AB$ collisions,  
the peak is shifted by about 0.4 GeV in the direction of
large transverse momenta. On the other hand, 
the contributions from the scale $\mu \sim p_T$ 
to $M$ have been resummed. Resummation plays a
more important role for small $p_T$ than for large $p_T$.
In EKS98, the strict shadowing effect is more important for small scales,
such as $p_T \sim 2-5$ GeV, than for  $p_T \sim M$. Thus, a relatively 
small shift in the peak position brings about the rise of $R_{sh}$ above 
unity. The shadowing ratio $R_{sh}$ dips below one again at a certain
$p_T$, to approach the fixed-order pQCD result when $p_T \sim M$.
The situation in Fig. 9 can be understood by noting that 
shadowing in the resummed transverse momentum distributions is
sensitive to the scale used to describe the nuclear effects on the 
parton distribution function.

After this discussion one may want to revisit Fig.~8 and 
ask why is $R_{sh}$ flat at RHIC energies. The answer lies 
in the observation that the EMC region of $x$ dominates the 
cross section at $\sqrt{s}=$ 350 MeV. There is no sensitivity 
to the scale in that region in EKS98.

In summary, the shadowing effects in the $p_T$ distribution of 
heavy bosons at RHIC and LHC are sensitive to the scale
of the nuclear parton distribution. The available data on the scale 
dependence of nuclear parton distribution functions 
are very limited\cite{eks}. Theoretical
studies (such as EKS98) are based on the the assumption 
that the nuclear parton distribution functions differ from the
parton distributions in the free proton, but obey the same DGLAP 
evolution\cite{eks}. Therefore, the tranverse momentum distribution 
of heavy bosons at LHC in $Pb + Pb$ collisions can provide a 
further test of our understanding of the nuclear parton distributions.

\section{Enhanced power corrections in nuclear collisions}
\label{sec4}

We have seen in Section \ref{sec2} that the effect of power corrections on 
the resummed cross section is very weak in $pp$ collisions
at both RHIC and LHC.
In $pA$ and $AB$ collisions, power corrections will be enhanced due
to rescattering in the nucleus. In this Section we discuss 
the enhancement of power corrections in nuclear collisions.

As ${\bar g}_2$ represents the effect of the partons' non-vanishing intrinsic
transverse momentum, it should not have a strong nuclear dependence. On the
other hand, $g_2$ arises as a result of 
dynamical power corrections, and should be enhanced by the nuclear
size, i.e. proportional to $A^{1/3}$. Considering the enhancement,
we get $G_2\sim 0.8$ GeV$^2$ for $p + Au$ and $G_2 \sim 1.5$ GeV$^2$ for $Pb + Pb$ reactions.

In order to study the effects of power corrections,
we now define $R_{G_2}^{sh}$ in nuclear collision as follows:
\begin{equation}
R_{G_2}^{sh}(p_T) \equiv \left.
\frac{d\sigma^{(G_2,sh)}(p_T,Z_A/A, Z_B/B)}
{dp_T} \right/
\frac{d\sigma^{(sh)}(p_T,Z_A/A, Z_B/B)}{dp_T} \,\,\, ,
\label{rg2-sh}
\end{equation}
 where the numerator represents the result of the full calculation,
with isospin effects, shadowing, and power corrections taken into account,  
and the denominator was introduced in connection with Eq. (\ref{Sigma-sh})
and contains no power corrections ($G_2 = 0$).

Fig.10(a) and Fig.11(a) present the $ d\sigma^{(G_2,sh)}(p_T,Z_A/A, Z_B/B)$
for $W^\pm$ production at $\sqrt{s}=350$ GeV in $p + Au$ reactions 
and $\sqrt{s}=5.5$ TeV in $Pb + Pb$ collisions, respectively.
In Fig. 10(b) the solid line is
$R_{G_2}^{sh}$ in $p + Au$ collisions and the dashed line is 
$R_{G_2}$  in $pp$ collisions at the same energy. At $p_T\sim 0$ GeV,
$R_{G_2}^{sh}$ is about 0.88, in contrast to about 0.95 in $pp$
collisions. The effects of power corrections are enhanced 
about three times in $p+ Au$ collision relative to $pp$ collisions. 
However, even for $p_T \sim 0$, the effect of power corrections 
is only about 10\%, and for $p_T > 2$ GeV, the effect remains under 5\%. 
Thus the effect of power corrections is weak in $p + Au$ collisions at RHIC. 
As for the situation at LHC in $Pb + Pb$ collisions, displayed in Fig. 11, 
the effects of power corrections appear to be enhanced by a similar factor 
(about three) from $pp$ to $Pb + Pb$ collisions at LHC.  The reason for
the similar enhancement is the interplay of the higher energy and the 
larger value of $G_2$ mentioned above. In any case,
the enhanced power corrections remain under 3\% in the [0,80] GeV 
$p_T$ interval for $Pb + Pb$ at LHC. The results for $Z^0$ production 
are similar to those for $W^\pm$ production. 

\section {Discussion and conclusion}
\label{sec5}

We have demonstrated that the predictive power of a
resummed pQCD calculation for heavy boson production is very good 
at RHIC and excellent at LHC, because the dependence of the 
cross sections on the non-perturbative input is weak. This makes 
$W^\pm$ and $Z^0$ production reliably calculable at the energies
considered. On the experimental side, $W^\pm$ and $Z^0$ production
should be an early observable at LHC (with sufficiently 
high production rates) in both $pp$ and nuclear collisions.
At RHIC, production rates may remain on the low side, in particular
in $pp$ collisions. In all cases, the cross sections peak at 
relatively low transverse momenta, and we calculate the peak 
position for all systems considered. 

We have examined the effects of the isospin composition of 
nuclei and of shadowing in nuclear collisions. We find that isospin effects 
are modest in $W^\pm$ and $Z^0$ production at the energies considered,
while shadowing is relatively unimportant at RHIC, but plays an 
important role at LHC in shaping the transverse momentum distributions.
This comes about because of the sensitivity of shadowing effects to
the scale of the parton distribution functions. We also
studied the magnitude of power corrections in $pp$ and nuclear 
collisions. Power corrections are small in $pp$ collisions, and,
even though part of these corrections is enhanced in nuclear 
collisions by a factor of $A^{1/3}$, they remain below 10\% 
(even at very low transverse momenta) in $AB$ reactions. 

As indicated by the above observations, the resummed pQCD formalism
yields accurate results for $W^\pm$ and $Z^0$ transverse momentum
distributions at RHIC and LHC. The production of heavy bosons can therefore 
serve as a further test of the nuclear parton distributions. In more 
general terms, $W^\pm$ and $Z^0$ production at RHIC and LHC 
provides a bench mark test of resummed pQCD for $pp$ and
nuclear collisions at these energies. We 
hope that our predictions will soon be compared to data.  


\section*{Acknowledgments}

We thank T. Hallman, D. Keane, and R. Vogt for stimulating discussions
and J.W. Qiu for a careful reading of the manuscript.
This work was supported in part by the U.S. Department of Energy under
Grant No. DE-FG02-86ER-40251 and by the NSF under Grant INT-0000211.


\section{Appendix}
\label{secapp}

Here we summarize relevant details of the CSS formalism concerning the
structure of the terms in Eq.~(\ref{css-gen}). We address,
in particular, the construction of the exponential factor $\exp[S(\tb,M)]$
in Eq.~(\ref{css-W-sol}).

It was shown in Ref. \cite{CSS-W}
that for $\tb\ll 1/\Lambda_{\rm QCD}$, the function
$\tilde{W}(\tb,M,x_A,x_B)$ of Eq.~(\ref{css-gen})
is directly related to the singular parts of
the  $p_T$ distribution as $p_T \rightarrow 0$. More
precisely, $\tilde{W}(\tb,M,x_A,x_B)$ includes all singular terms
like $\delta^2(\vec{p}_T)$ and $[\ln^m(p^2/p_T^2)/p_T^2]_{\rm reg}$
with $m \ge 0$. Less singular terms as
$p_T\rightarrow 0$ are included in the $Y$ term of
Eq.~(\ref{css-gen}). The QCD resummation of the large logarithms in
the CSS formalism is achieved by solving the evolution equation
for the $\tilde{W}_{ij}$ of Eq.~(\ref{css-W-ij}),
\begin{equation}
\frac{\partial}{\partial\ln M^2} \tilde{W}_{ij}(\tb,M,x_A,x_B)
= \left[ K(\tb\mu,\alpha_s(\mu)) + G(M/\mu,\alpha_s(\mu)) \right]
\tilde{W}_{ij}(\tb,M,x_A,x_B) \,\,\, ,
\label{css-W-evo}
\end{equation}
and the corresponding renormalization group equations for the kernels $K$
and $G$,
\begin{eqnarray}
\frac{\partial}{\partial\ln\mu^2} K(\tb\mu,\alpha_s(\mu))
&=& -\frac{1}{2} \gamma_K(\alpha_s(\mu)) \,\, ,
\label{css-K-rg} \\
\frac{\partial}{\partial\ln\mu^2} G(M/\mu,\alpha_s(\mu))
&=& \frac{1}{2} \gamma_K(\alpha_s(\mu)) \,\, .
\label{css-G-rg}
\end{eqnarray}
The anomalous dimension $\gamma_K(\alpha_s(\mu))=\sum_{n=1}
\gamma_K^{(n)} (\alpha_s(\mu)/\pi)^n$ in Eqs.~(\ref{css-K-rg}) and
(\ref{css-G-rg}) is perturbatively calculable \cite{CSS-W}. The
renormalization group equations (\ref{css-K-rg}) and (\ref{css-G-rg})
for $K$ and $G$ ensure the correct
renormalization scale dependence,
\begin{equation}
\frac{d}{d\ln\mu^2} \tilde{W}(\tb,M,x_A,x_B) = 0 \,\, .
\end{equation}
The solution given in
Eq.~(\ref{css-W-sol}) corresponds to simultaneously
solving the evolution equation (\ref{css-W-evo}) from
$\ln(c^2/\tb^2)$ to $\ln(M^2)$, and the renormalization group
equations (\ref{css-K-rg}) and (\ref{css-G-rg}) from
$\ln(c^2/\tb^2)$ to $\ln(\mu^2)$ and from $\ln(M^2)$ to $\ln(\mu^2)$,
respectively,  where $c$ is a constant of order unity.

Integrating Eq.~(\ref{css-K-rg}) 
from $\ln(c^2/\tb^2)$
to $\ln(\mu^2)$, and Eq.~(\ref{css-G-rg}) from $\ln(M^2)$ to
$\ln(\mu^2)$, one derives
\begin{equation}
K(\tb\mu,\alpha_s(\mu)) + G(M/\mu,\alpha_s(\mu))
= -\int_{c^2/\tb^2}^{M^2}\,
\frac{d\bar{\mu}^2}{\bar{\mu}^2}\,
A(\alpha_s(\bar{\mu}))
- B(\alpha_s(M)) \,\, ,
\label{css-KG}
\end{equation}
where $A$ is a function of $\gamma_K(\alpha_s(\bar{\mu}))$ and
$K(c,\alpha_s(\bar{\mu}))$ while $B$ depends on both
$K(c,\alpha_s(M))$ and $G(1,\alpha_s(M))$.
The functions $A$ and $B$ do not have large logarithms and have
perturbative expansions $A=\sum_{n=1} A^{(n)}(\alpha_s/\pi)^n$ and
$B=\sum_{n=1} B^{(n)}(\alpha_s/\pi)^n$, respectively.
The first two coefficients in the perturbative expansions are
known\cite{CSS-W,Davis}:
\begin{eqnarray}
A^{(1)} &=& C_F\ ,
\nonumber\\
A^{(2)} &=& \frac{C_F}{2} \left[
N\left(\frac{67}{18}-\frac{\pi^2}{6}\right)
-\frac{10}{9}T_R\, n_f \right]\, ,
\nonumber\\
B^{(1)} &=& -\frac{3}{2}C_F\, ,
\nonumber\\
B^{(2)} &=& \left(\frac{C_F}{2}\right)^2
\left[\pi^2 -\frac{3}{4}-12\zeta(3)\right]
+ \frac{C_F}{2}\, N
\left[\frac{11}{18}\pi^2-\frac{193}{24} + 3\zeta(3)
\right]
\nonumber\\
&+& \frac{C_F}{2}T_R\, n_f
\left[\frac{17}{6}-\frac{2}{9}\pi^2\right]\, ,
\label{css-AB}
\end{eqnarray}
where $N=3$ for SU(3) color, the color factor $C_F=(N^2-1)/2N=4/3$
for quarks, $T_R=1/2$, and
$n_F$ is the number of active quark flavors. The coefficients $A$ and
$B$ given in Eq.~(\ref{css-AB}) are derived from the general
expressions in Ref.~\cite{CSS-W} with 
renormalization constant $c=2{\rm e}^{-\gamma_E}$,
where $\gamma_E\approx 0.577$ is Euler's constant\cite{CS-b}.

Substituting Eq.~(\ref{css-KG}) into Eq.~(\ref{css-W-evo}), and
integrating 
from $\ln(c^2/\tb^2)$ to $\ln(M^2)$, one
obtains the $\tilde{W}_{ij}$ given in Eq.~(\ref{css-W-sol}) with
\begin{equation}
S(\tb,M) = -\int_{c^2/\tb^2}^{M^2}\,
\frac{d\bar{\mu}^2}{\bar{\mu}^2} \left[
\ln\left(\frac{M^2}{\bar{\mu}^2}\right)
A(\alpha_s(\bar{\mu})) + B(\alpha_s(\bar{\mu})) \right]\, .
\label{css-S}
\end{equation}
In Eq.~(\ref{css-W-sol}), all large logarithms from $\ln(c^2/\tb^2)$ to
$\ln(M^2)$ in $\tilde{W}_{ij}(\tb,M,x_A,x_B)$ are completely resummed
into the exponential factor $\exp[S(\tb,M)]$, leaving the
$\tilde{w}_{ij}(\tb,c/\tb,x_A,x_B)$ with only one momentum scale
$1/\tb$. The $\tilde{w}_{ij}(\tb,c/\tb,x_A,x_B)$ in
Eq.~(\ref{css-W-sol}) is then perturbatively calculable when
the momentum scale $1/\tb$ is large enough, and is given by
\cite{CSS-W,Ellis-1}
\begin{equation}
\tilde{w}_{ij}(\tb,c/\tb,x_A,x_B) =
f_{i/A}(x_A,\mu=c/\tb)\, f_{j/B}(x_B,\mu=c/\tb)\, .
\label{css-W-pert}
\end{equation}
The functions $f_{i/A}$ and $f_{j/B}$ are the modified parton
distributions \cite{CSS-W,Ellis-1},
\begin{equation}
f_{i/A}(x_A,\mu) = \sum_a
\int_{x_A}^1\frac{d\xi}{\xi}\,
C_{i/a}(x_A/\xi,\mu))\, \phi_{a/A}(\xi,\mu) \,\, ,
\label{mod-pdf}
\end{equation}
where $\sum_a$ runs over all parton flavors. In Eq.~(\ref{mod-pdf}),
$\phi_{a/A}(\xi,\mu)$ is the normal parton distribution for finding a
parton of flavor $a$ in hadron $A$,
and $C_{i/a}=\sum_{n=0} C_{i/a}^{(n)} (\alpha_s/\pi)^n$
are perturbatively calculable coefficient functions for finding a
parton $i$ in the modified parton distribution $f_{i/A}(x_A,\mu)$
from a parton $a$ in the normal parton distribution $\phi_{a/A}(\xi,\mu)$.
The first two coefficients of the
$C_{i/a}$ are available\cite{CSS-W,Davis}:
\begin{eqnarray}
C^{(0)}_{i/j}(z,\mu=c/\tb) &=& \delta_{ij}\, \delta(z-1)\, ,
\nonumber\\
C^{(0)}_{i/g}(z,\mu=c/\tb) &=& 0\, ,
\nonumber \\
C^{(1)}_{i/j}(z,\mu=c/\tb) &=& \delta_{ij} \frac{C_F}{2} \left[
(1-z)
+ \left( \frac{\pi^2}{2}-4 \right) \delta(1-z) \right] ,
\nonumber \\
C^{(1)}_{i/g}(z,\mu=c/\tb) &=& T_R\, z\, (1-z)  \,\,\, ,
\label{css-coef}
\end{eqnarray}
where $i$ and $j$ represent quark or antiquark flavors and $g$
represents a gluon. The coefficient functions given in
Eq.~(\ref{css-coef}) are derived from the general functional forms in
Ref.~\cite{CSS-W} by setting the renormalization constants and the
factorization scale as, $C_1=c$, $C_2=1$, and $\mu=c/\tb$.

The $\sigma_{ij\rightarrow V}(M)$ in Eq.~(\ref{css-W-ij}) is the
lowest order cross section for a quark-antiquark pair to
annihilate into the vector boson ($V=\gamma^*$, $W^{\pm},$ or $Z^0$).
For $V=Z^0$, we have\cite{CS-b}
\begin{equation}
\sigma_{ij\rightarrow Z^0}(M_Z) =
Q_{ij}
\left(\frac{4\pi^3\alpha^2_{EM}(M_z)}{3 s}\right)  \,\,\, ,
\label{css-W-lo}
\end{equation}
where $Q_{ij}$ is the  weak charge,
\begin{equation}
Q_{ij}={[1-4|e_q|{\sin^2\theta_W}]^2+1
\over 16\sin^2\theta_W\cos^2\theta_W} \,\,\,  ,
\end{equation}
with $\theta_W$ denoting the Weinberg angle and $e_q$ denoting
the charge of the quark. Here, $i$ and $j$ must represent a 
quark-antiquark pair of the same flavor. For the Weinberg angle we use
$\sin^2\theta_W=0.23$.
The $\sigma_{ij\rightarrow V}(M)$ for $V=W^{\pm}$ and $\gamma^*$ 
can be found in Refs.~\cite{CSS-W,Ellis-1}.

In the CSS resummation formalism, the $Y$ term in Eq.~(\ref{css-gen})
represents a small correction to the $p_T$ distribution when $p_T \ll M$.
However, the $Y$ term dominates the $p_T$-distributions when $p_T \sim M$.
The $Y$ term has the perturbative
expansion $Y=\sum_{n=1} Y^{(n)} (\alpha_s(\mu)/\pi)^n$, and the
coefficients $Y^{(n)}$ have the factorized form\cite{CSS-W}
\begin{eqnarray}
Y^{(n)}(p_T,M,x_A,x_B;\mu) &=& \sum_{a,b}
\int_{x_A}^1 \frac{d\xi_A}{\xi_A}\, \phi_{a/A}(\xi_A,\mu)
\int_{x_B}^1 \frac{d\xi_B}{\xi_B}\, \phi_{b/B}(\xi_B,\mu)
\nonumber\\
&\times &
\left(\frac{4\pi^3\alpha^2_{EM}(M_z)}{3 s}\right) \ 
R_{ab\rightarrow V}^{(n)}(p_T,M,x_A/\xi_A,x_B/\xi_B;\mu) \,\,\, ,
\label{css-W-yn}
\end{eqnarray}
where $\sum_{a,b}$ runs over all possible parton flavors and $\mu$
represents both the factorization and renormalization scales.
The $R_{ab\rightarrow V}^{(n)}$ in Eq.~(\ref{css-W-yn})
are perturbatively calculable and have the same normalization as those
introduced in Ref.~\cite{CSS-W}.

Since the perturbatively resummed $\tilde{W}_{ij}(\tb,M,x_A,x_B)$ in
Eq.~(\ref{css-W-sol}) is only reliable for the small $\tb$ region, an
extrapolation to the large $\tb$ region is necessary in order to
complete the Fourier transform in Eq.~(\ref{css-gen}).
In the original CSS formalism, a variable $\tb_*$
and a nonperturbative function $F_{CSS}^{NP}(\tb,M,x_A,x_B)$
were introduced\cite{CSS-W}  in such a way that $\tilde{W}$
was modified for all $\tb$ (except $\tb=0$) as
\begin{equation}
\tilde{W}^{\rm CSS}(\tb,M,x_A,x_B) \equiv
\tilde{W}(\tb_*,M,x_A,x_B)\,
F_{CSS}^{NP}(\tb,M,x_A,x_B)\, ,
\label{css-W-b}
\end{equation}
where $\tb_*=\tb/\sqrt{1+(\tb/\tb_{max})^2}$, with 
$\tb_{max} = 0.5$ GeV$^{-1}$.  This construction ensures that 
$\tb_* \leq \tb_{max}$ for all values of $\tb$. The function
$F_{CSS}^{NP}$ is a Gaussian in $\tb$, $F_{CSS}^{NP} \sim
\exp(-\kappa \tb^2)$ with $\kappa$ having a certain
$M^2$, $x_A$, and $x_B$ dependence\cite{CSS-W}. 

As it has been pointed out in this paper and earlier\cite{qz01},
the original CSS extrapolation introduces a
modification to the perturbative calculation, and the size of the
modifications depends on the non-perturbative parameters in
$F_{CSS}^{NP}(\tb,M,x_A,x_B)$.
 This motivated the development of the alternative extrapolation
formula\cite{qz01} used in the present work.


\begin{figure}
\begin{center}
\epsfig{figure=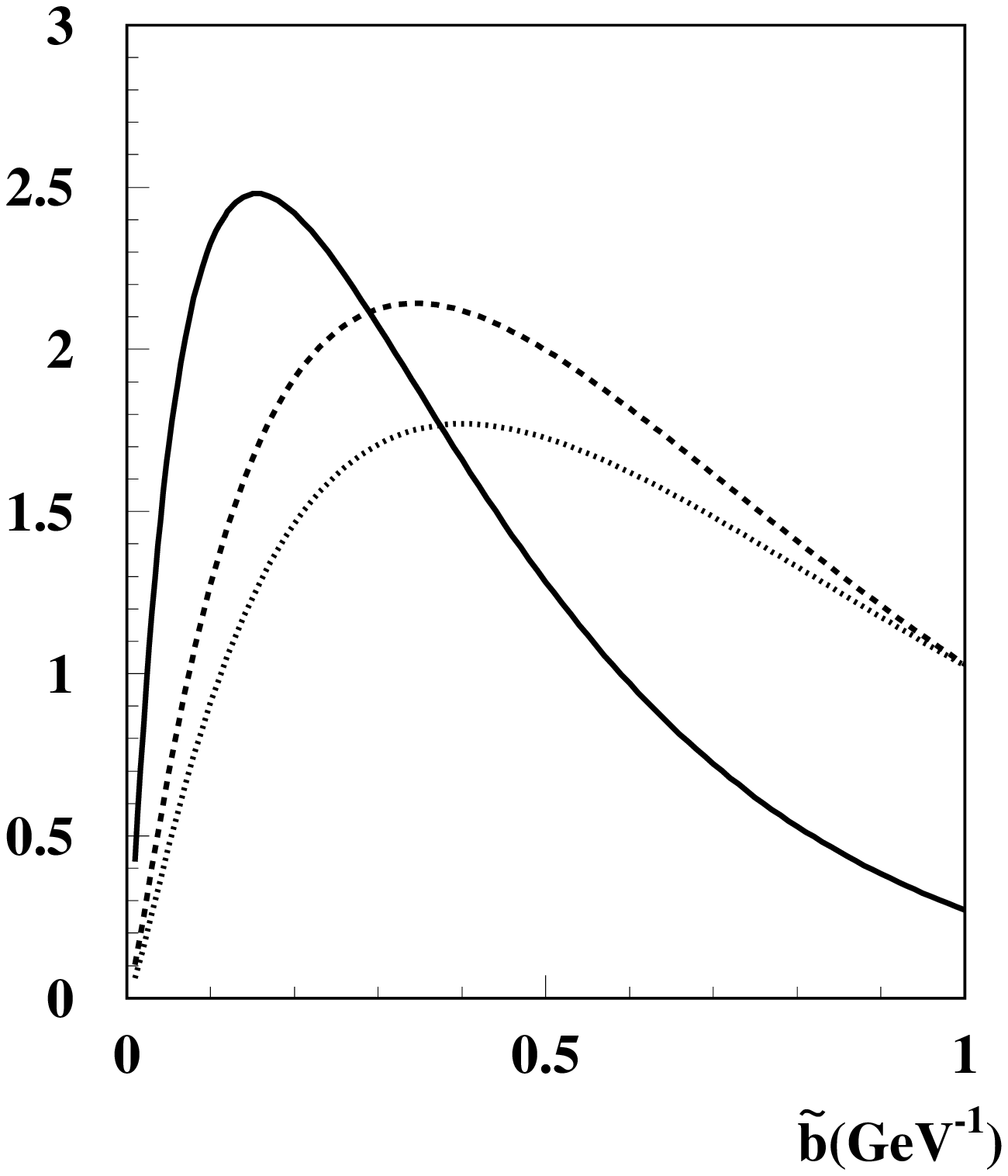,width=4.5in}
\end{center}
\caption{Integrand of the $\tb$-integration in
Eq.~(\protect\ref{css-W-F}) at $p_T=0$ for $Z^0$ as a function of
$\tb$ with an arbitrary normalization at $\sqrt{s}=350$ GeV (dotted line),
$\sqrt{s}=500$ GeV (dashed), and $\sqrt{s}=5.5$~TeV (solid).}  
\label{fig1}
\end{figure} 

\begin{figure}
\begin{center}
\epsfig{figure=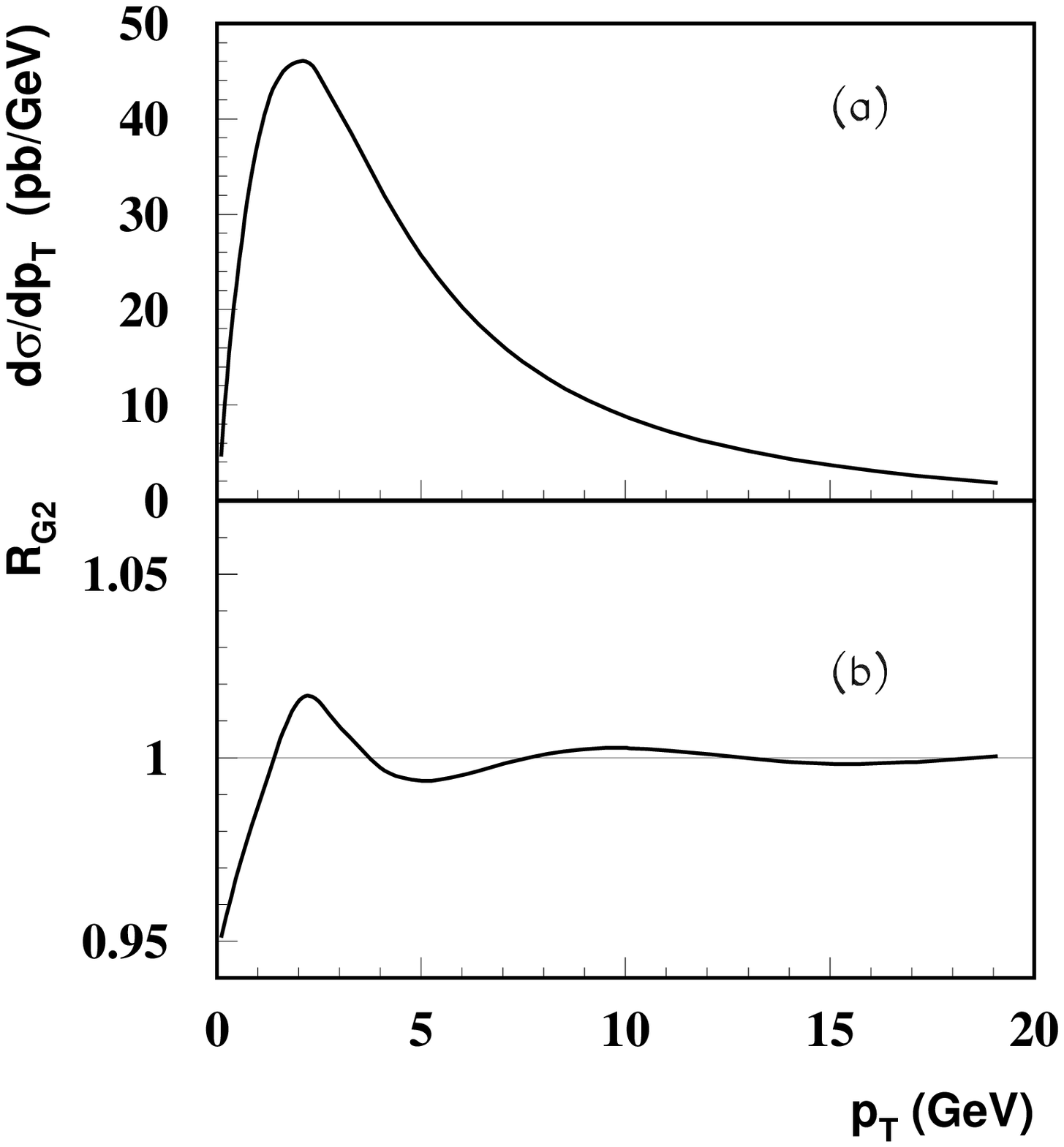,width=5.5in}
\end{center}
\caption{(a) Cross section ${d \sigma \over dp_T}$ for $Z^0$ production 
in $pp$ collisions at RHIC with
$\sqrt{s}=500$ GeV; (b) $R_{G_2}$ defined in Eq.~(\protect\ref{Sigma-g2})
for $Z^0$ at the same energy.}
\label{fig2}
\end{figure} 

\begin{figure}
\begin{center}
\epsfig{figure=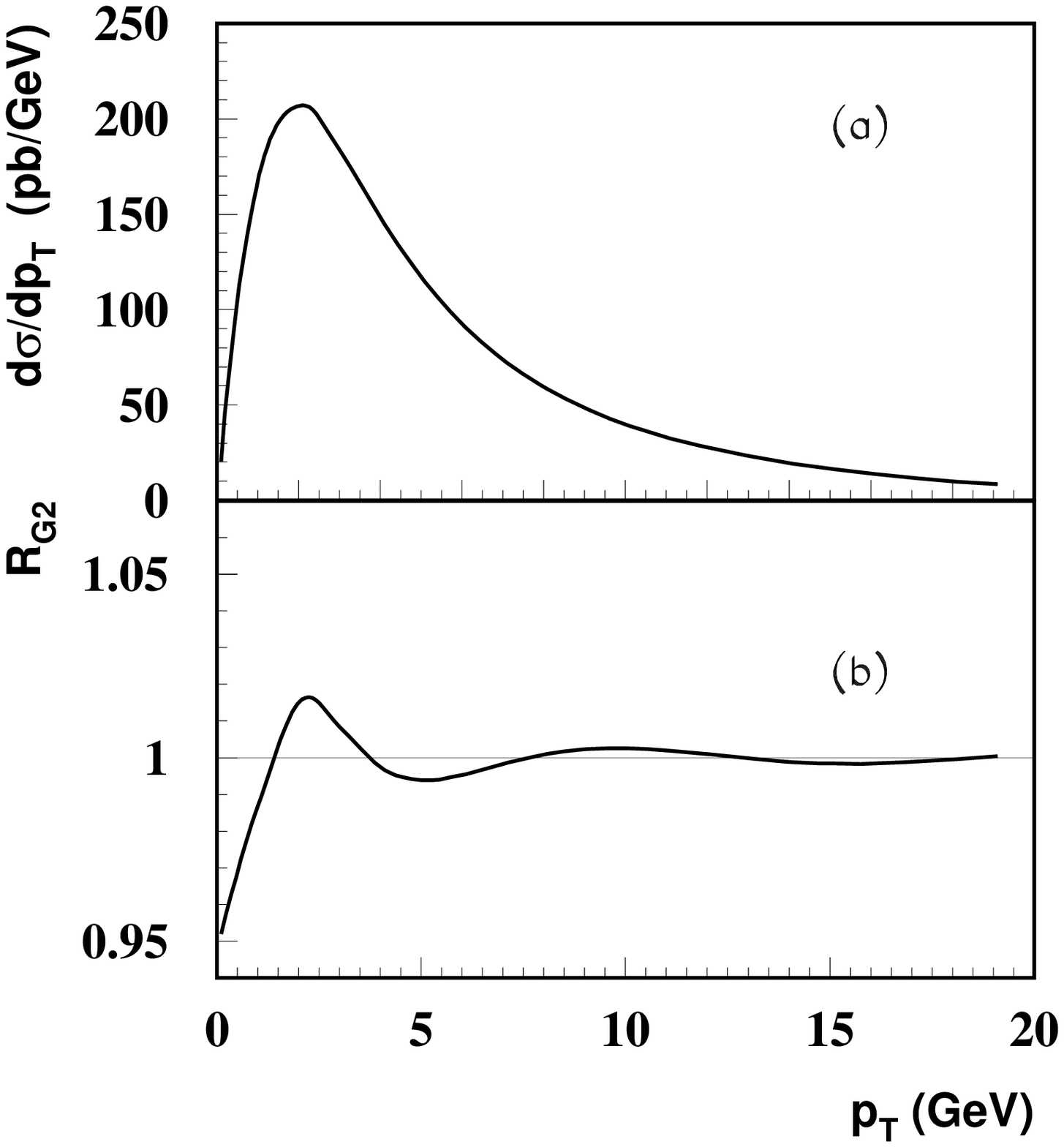,width=5.5in}
\end{center}
\caption{(a) Cross section ${d \sigma \over dp_T}$ for $W^\pm$ production in 
$pp$ collisions at RHIC with
$\sqrt{s}=500$ GeV; (b) $R_{G_2}$ defined in Eq.~(\protect\ref{Sigma-g2})
for $W^\pm$ at the same energy.}
\label{fig3}
\end{figure} 

\begin{figure}
\begin{center}
\epsfig{figure=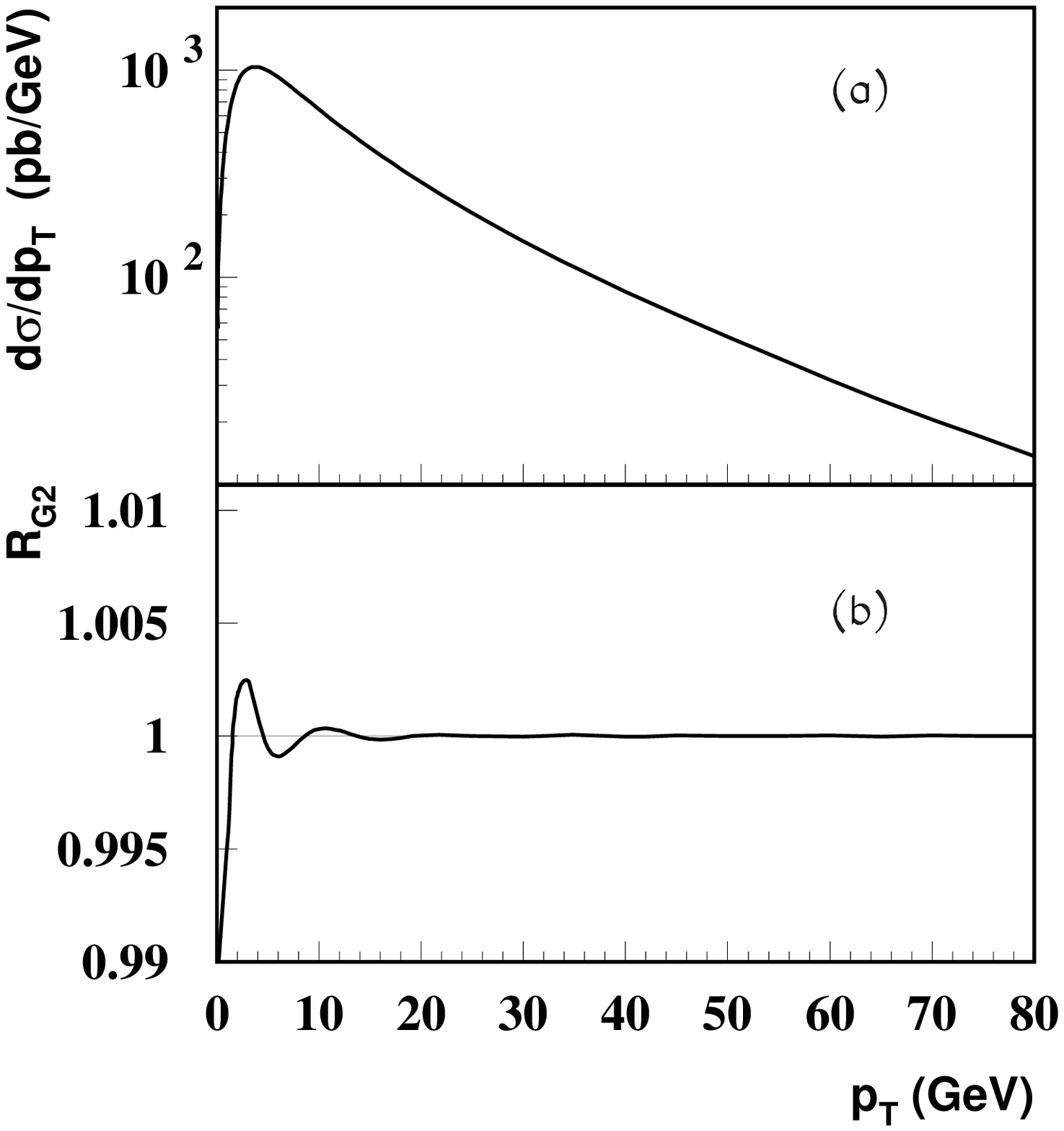,width=5.5in}
\end{center}
\caption{(a) Cross section ${d \sigma \over dp_T}$ for $Z^0$ production 
in $pp$ collisions at LHC with
$\sqrt{s}=5.5$ TeV; (b) $R_{G_2}$ defined in Eq.~(\protect\ref{Sigma-g2})
for $Z^0$ at the same energy.}
\label{fig4}
\end{figure} 

\begin{figure}
\begin{center}
\epsfig{figure=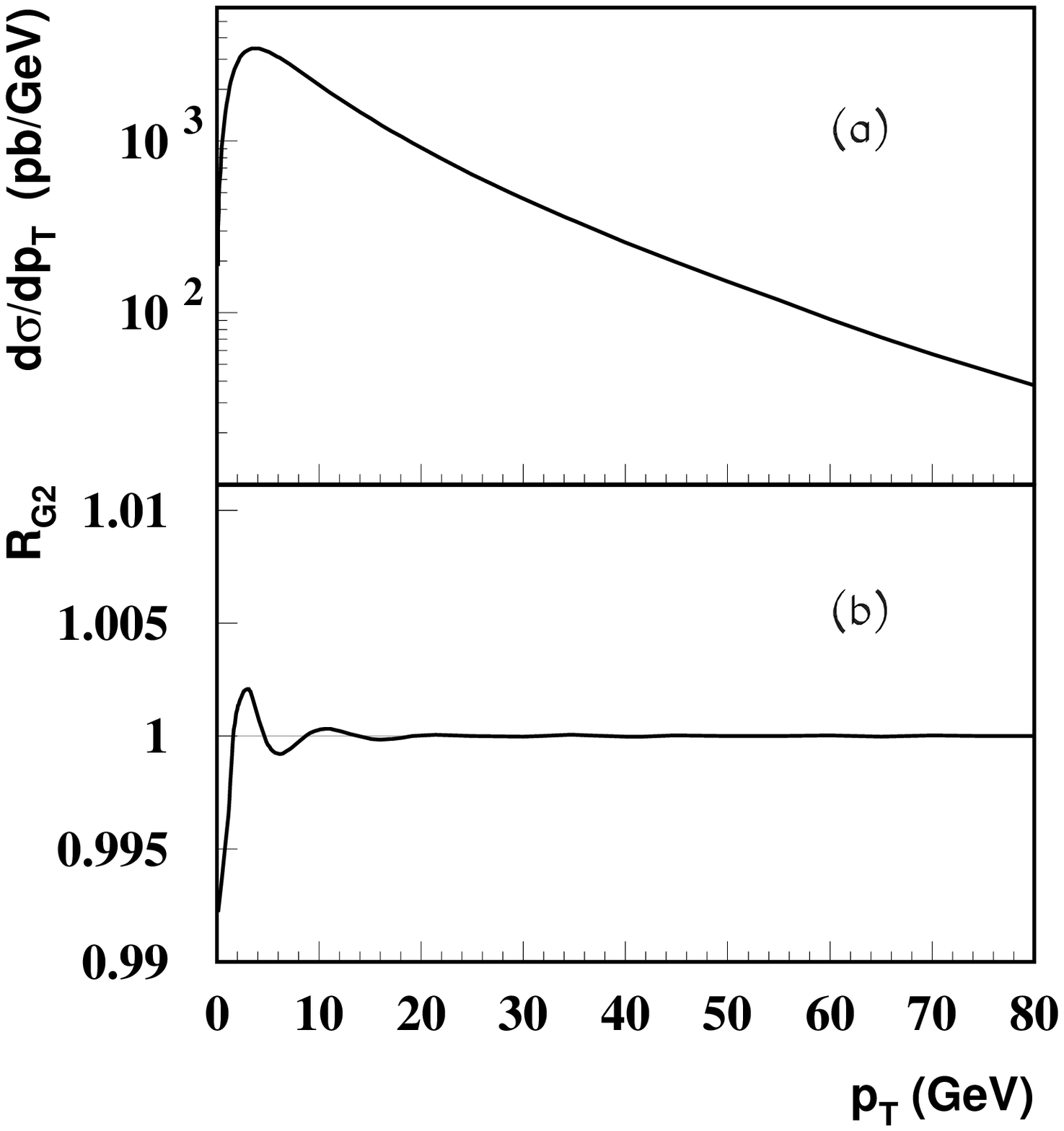,width=5.5in}
\end{center}
\caption{(a) Cross section ${d \sigma \over dp_T}$ for $W^\pm$ production 
in $pp$ collisions at LHC with
$\sqrt{s}=5.5$ TeV; (b) $R_{G_2}$ defined in Eq.~(\protect\ref{Sigma-g2})
for $W^\pm$ at the same energy.}
\label{fig5}
\end{figure}

\begin{figure}
\begin{center}
\epsfig{figure=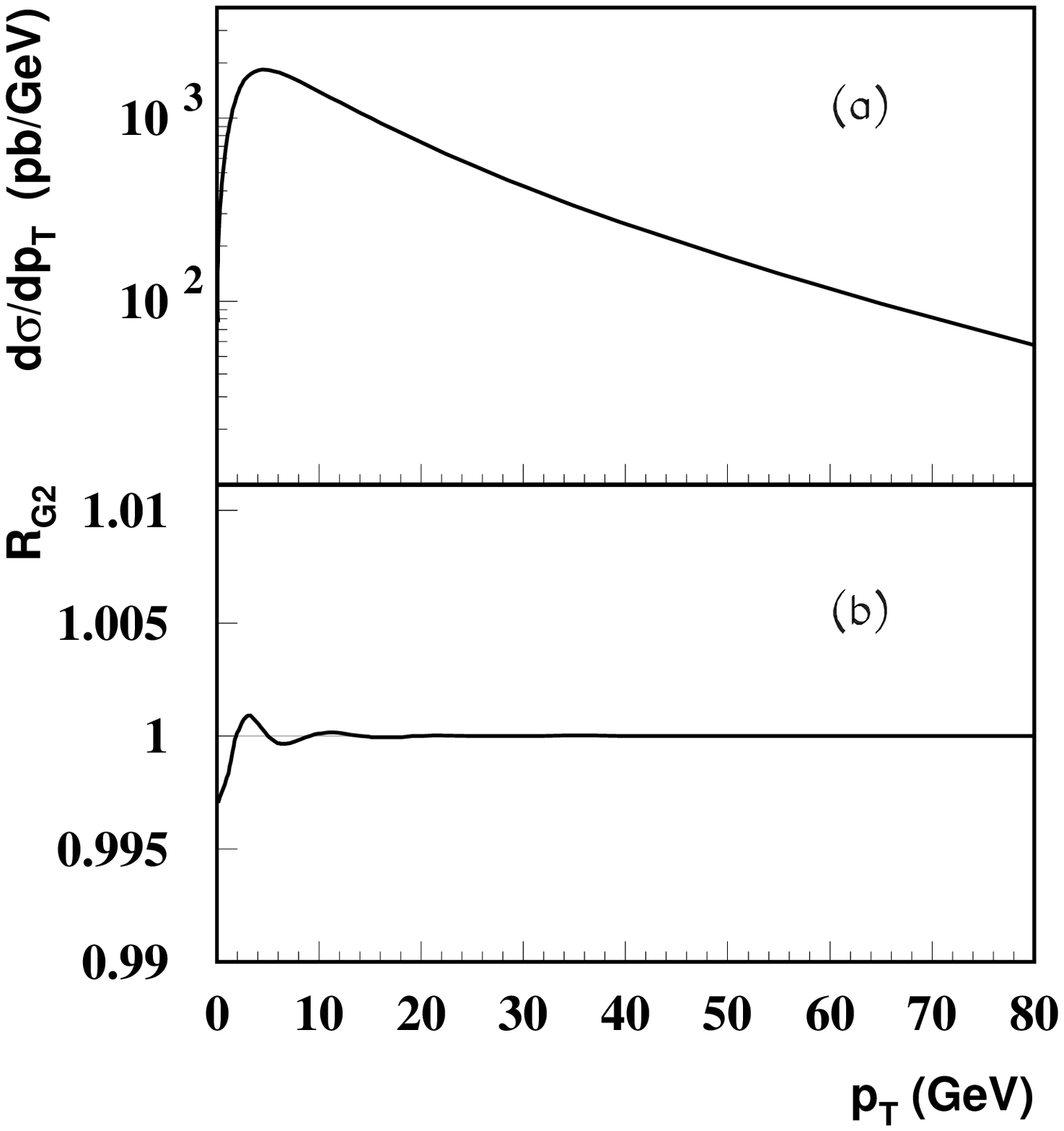,width=5.5in}
\end{center}
\caption{(a) Cross section ${d \sigma \over dp_T}$ for $Z^0$ production 
in $pp$ collisions at LHC with
$\sqrt{s}=14$ TeV; (b) $R_{G_2}$ defined in Eq.~(\protect\ref{Sigma-g2})
for $Z^0$ at the same energy.}
\label{fig6}
\end{figure}

\begin{figure}
\begin{center}
\epsfig{figure=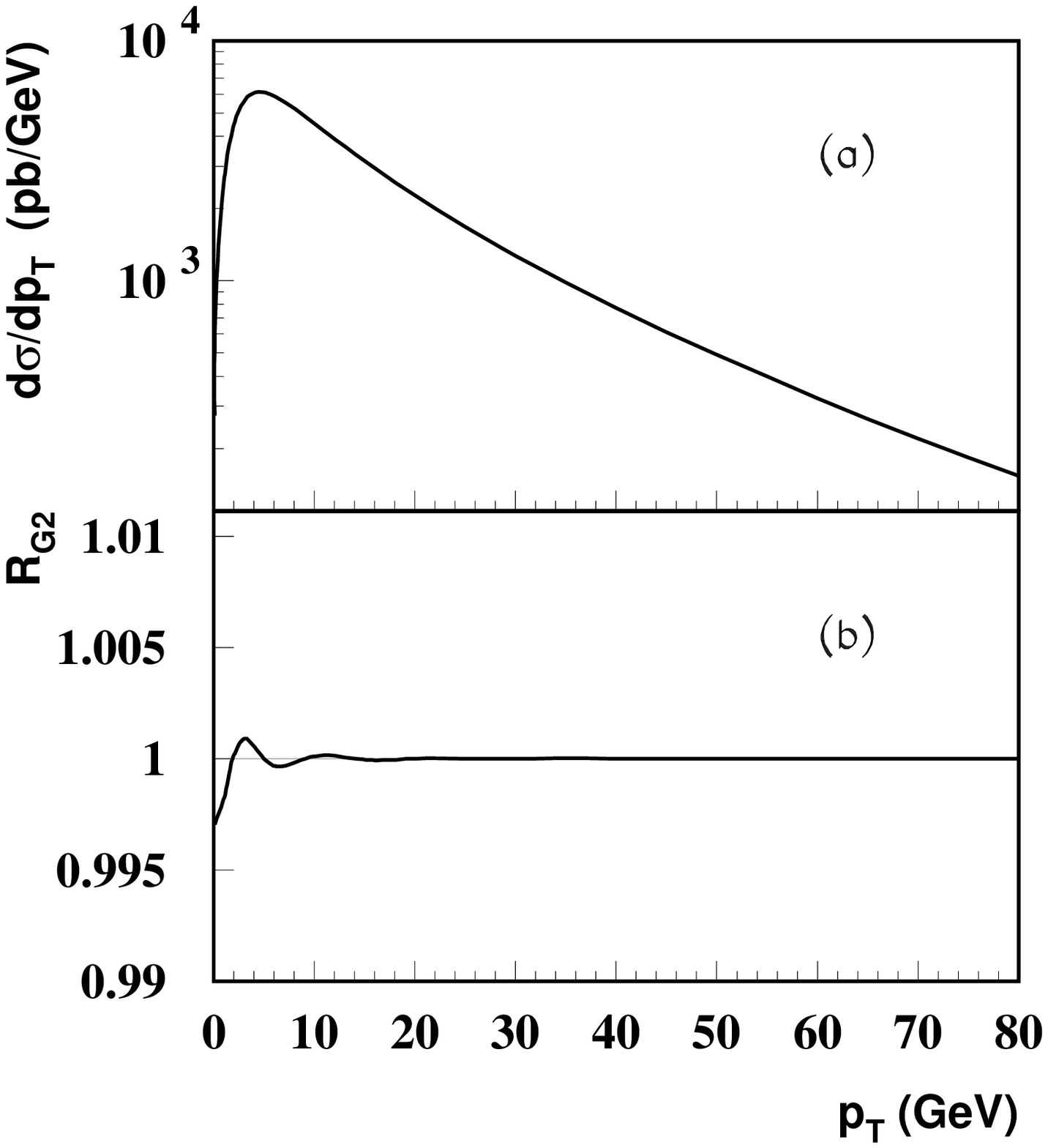,width=5.5in}
\end{center}
\caption{(a) Cross section ${d \sigma \over dp_T}$ for $W^\pm$ production 
in $pp$ collisions at LHC with
$\sqrt{s}=14$ TeV; (b) $R_{G_2}$ defined in Eq.~(\protect\ref{Sigma-g2})
for $W^\pm$ at the same energy.}
\label{fig7}
\end{figure}

\begin{figure}
\begin{center}
\epsfig{figure=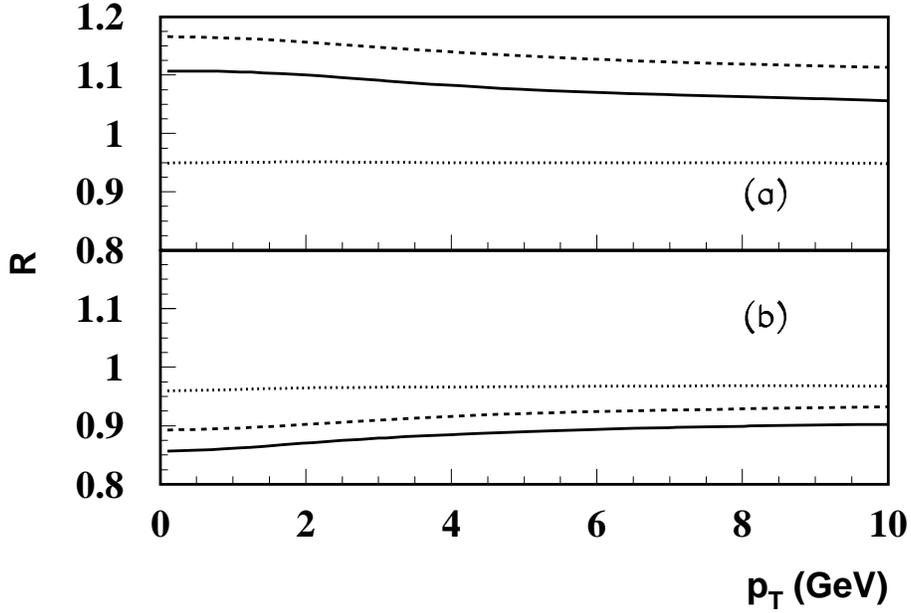,width=5.2in}
\end{center}
\caption{(a) The isospin ratio, $R_{iso}$, defined in 
Eq.~(\protect\ref{sigma_iso}) (dashed line),
the shadowing ratio, $R_{sh}$, 
defined in Eq.~(\protect\ref{Sigma-sh}) (solid line),
and ``pure shadowing'' (dotted)
for $Z^0$ in $p + Au$ collisions at $\sqrt{s}=350$ GeV; 
(b) same as (a) for $W^{\pm}$ production.}
\label{fig8}
\end{figure} 

\begin{figure}
\begin{center}
\epsfig{figure=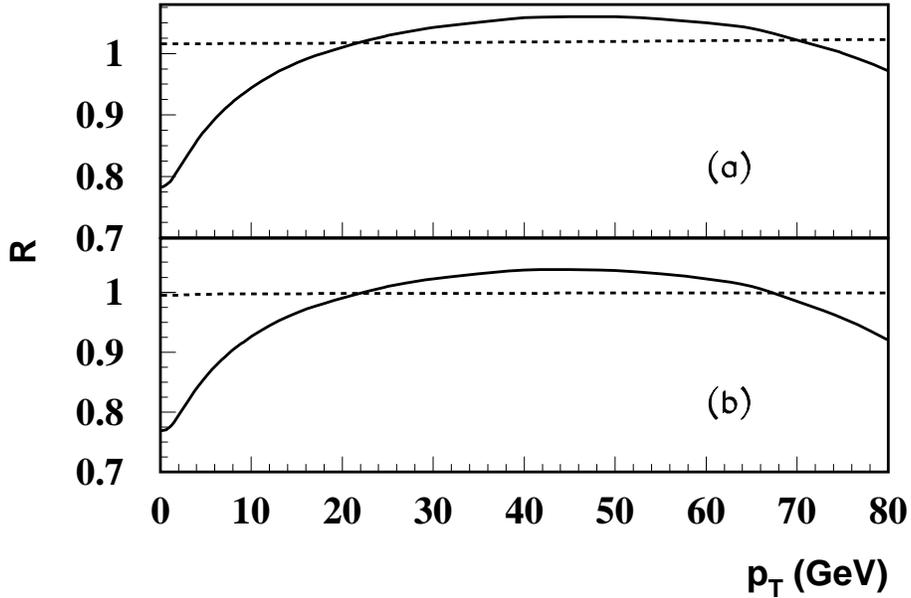,width=5.2in}
\end{center}
\caption{(a) The isospin ratio, $R_{iso}$ defined in 
Eq.~(\protect\ref{sigma_iso}) (dashed line) and
the shadowing ratio, $R_{sh}$, 
defined in Eq.~(\protect\ref{Sigma-sh}) (solid line)
for $Z^0$ in $Pb + Pb$ collisions at $\sqrt{s}=5.5$ TeV; 
(b) same as (a) for $W^{\pm}$ production.}
\label{fig9}
\end{figure}

\begin{figure}
\begin{center}
\epsfig{figure=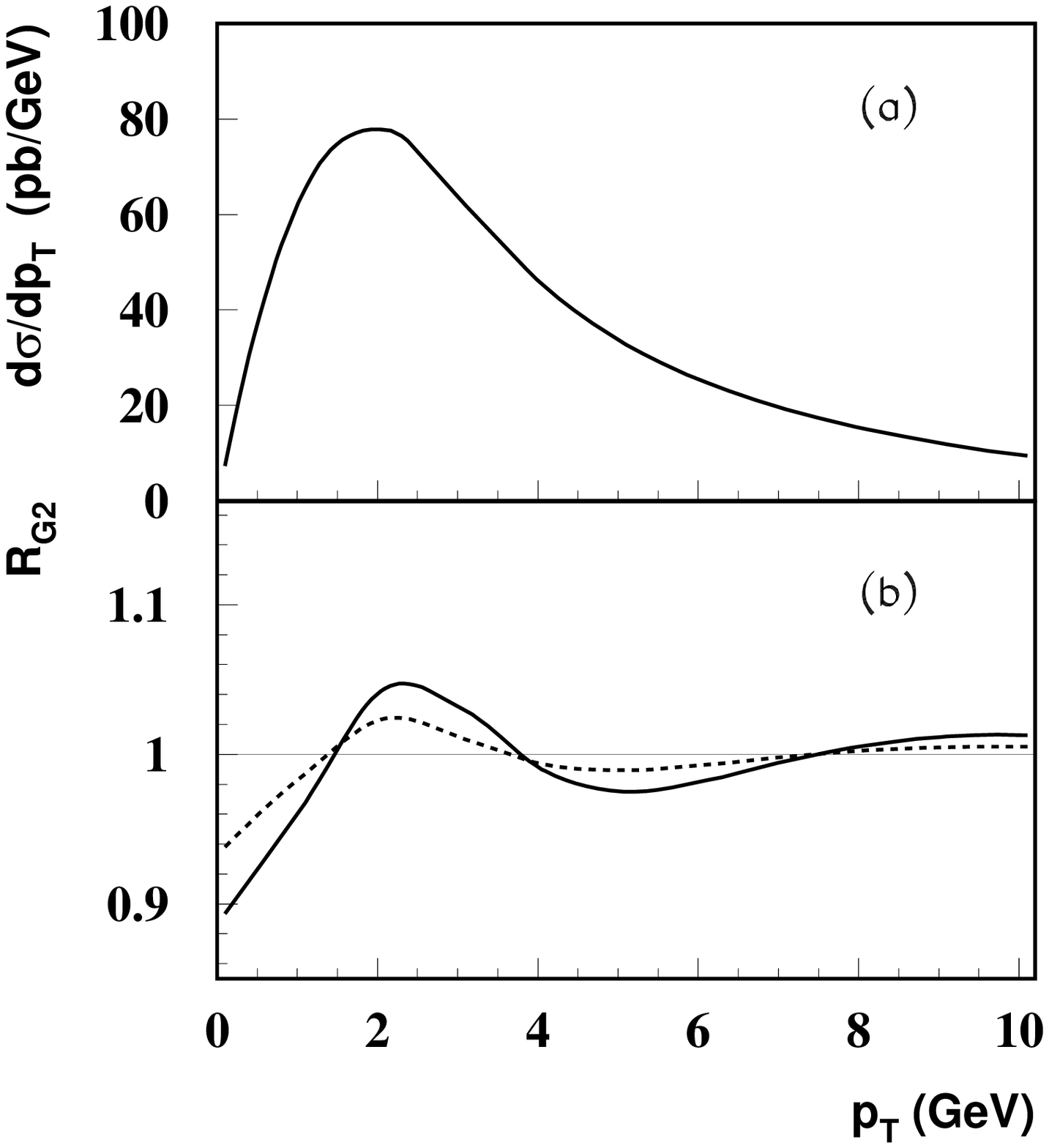,width=5.2in}
\end{center}
\caption{(a) Cross section ${d \sigma \over dp_T}^{(G_2,sh)}(p_T, Z_A/A,Z_B/B)$ 
for $W^\pm$ production in  $p + Au$ collisions 
at RHIC with $\sqrt{s}=350$ GeV averaged over $AB$;  
(b) The ratio $R_{G_2}^{sh}$ defined in Eq.~(\protect\ref{rg2-sh}) (solid line)
in $p + Au$ collision and $R_{G_2}$ defined in Eq.~(\protect\ref{Sigma-g2})
for $W^\pm$ at the same energy.}

\label{fig10}
\end{figure}  
\begin{figure}
\begin{center}
\epsfig{figure=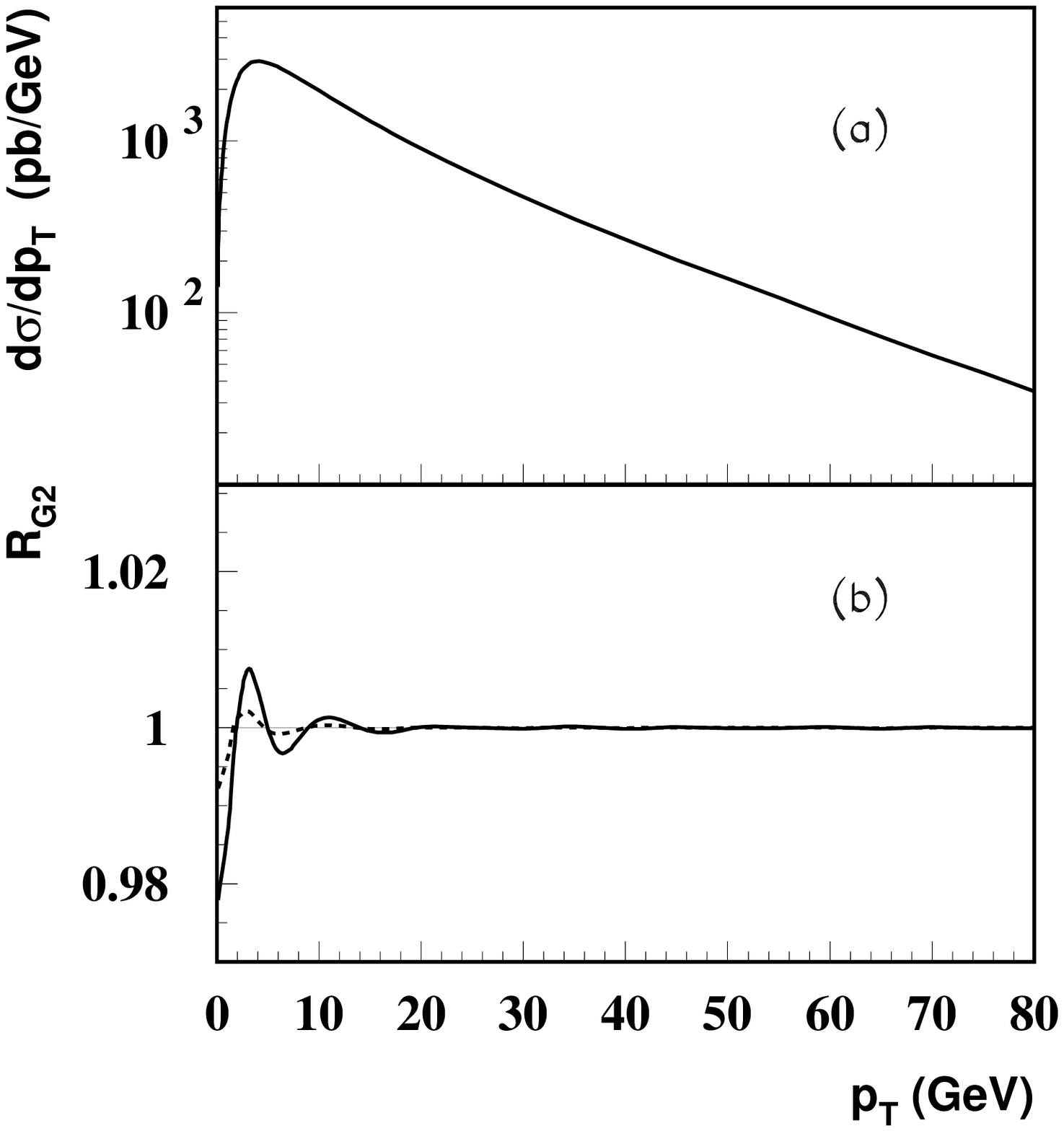,width=5.5in}
\end{center}
\caption{(a) Cross section ${d \sigma \over dp_T}^{(G_2,sh)}(p_T, Z_A/A,Z_B/B)$ 
for $W^\pm$ production in $Pb + Pb$ collisions 
at LHC with $\sqrt{s}=5.5$ TeV, averaged over $AB$;  
(b) The ratio $R_{G_2}^{sh}$ defined in Eq.~(\protect\ref{rg2-sh}) (solid line)
in $Pb + Pb$ collision and  $R_{G_2}$ defined in Eq.~(\protect\ref{Sigma-g2})
for $W^\pm$ at the same energy.}
\label{fig11}
\end{figure} 


\begin{references}

\bibitem{QM01}
Proceedings of {\it Quark Matter 2001}, Nucl. Phys. A {\bf 698}, 1 (2002);
http://www.rhic.bnl.gov/qm2001.

\bibitem{star}
C. Adler {\it et al.} (STAR) Phys. Rev. Lett. {\bf 87} 112303 (2001).

\bibitem{phenix1}
G. David {\it et al.} (PHENIX), Nucl. Phys. A {\bf 698}, 227 (2002),
nucl-ex/0105014; K. Adcox {\it et al.} (PHENIX), nucl-ex/0109003.

\bibitem{wang01}
E. Wang and X-N. Wang, Phys. Rev. Lett. {\bf 87}, 142301 (2001).

\bibitem{zfpbl02}
Y. Zhang, G. Fai, G. Papp, G.G. Barnafoldi, and P. Levai, Phys. Rev. C
{\bf 65}, 0349XX (2002).

\bibitem{CDF-Z}
T. Affolder {\it et al.}, CDF Collaboration, Phys. Rev. Lett.
{\bf 84}, 845 (2000).

\bibitem{D0-W}
B. Abbott {\it et al.}, D0 Collaboration, hep-ex/0010026.

\bibitem{CS-b}
J.C.~Collins and D.E.~Soper, Nucl. Phys. B {\bf 193}, 381 (1981);
Erratum {\bf 213}, 545 (1983); {\bf 197}, 446 (1982).

\bibitem{CSS-W}
J.C. Collins, D.E.~Soper and G.~Sterman, Nucl. Phys. B {\bf 250}, 199
(1985).

\bibitem{qz01}
J. Qiu and X. Zhang, Phys. Rev. Lett. {\bf 86}, 2724 (2001);
Phys. Rev. D {\bf 63}, 114011 (2001).

\bibitem{qiu-sterman-fac}
J. Qiu  and G. Sterman,hep-ph/0111002,
to appear in the Proceedings of the Hard Probe Collaboration. 

\bibitem{lq2}
J. Qiu and X. Zhang, hep-ph/0109210, to be appear in Phys. Lett. {\bf B}.

\bibitem{vogt01}
R. Vogt, Phys. Rev. {\bf C64} 044901(2001).

\bibitem{Ellis-1}
R.K.~Ellis, D.A.~Ross and S.~Veseli, Nucl.\ Phys.\ {\bf B503} 309
(1997).

\bibitem{Ellis-2}
R.K.~Ellis and S.~Veseli, Nucl.\ Phys.\ {\bf B511} 649 (1998).



\bibitem{AMS-DY}
D. Appell, G. Sterman, and P. Mackenzie, Nucl.Phys. {\bf B309}, 259 (1988).

\bibitem{PP-b}
G.~Parisi and R.~Petronzio, Nucl. Phys. {\bf B154}, 427 (1979).

\bibitem{zhang-fai}
X. Zhang and G. Fai, in preparation.

\bibitem{pinkbook}R. K. Ellis, W. J. Stirling, and B. R. Webber,
{\it QCD and Collider Physics},  Cambridge, England, 1996.

\bibitem{CTEQ5}H. L. Lai {et al.}, Eur. Phys. J. C {\bf 12}, 375( 2000).

\bibitem{wang1}
X.N. Wang and M. Gyulassy, Phys. Rev. D {\bf 44}, 3501 (1991).

\bibitem{wang2}
S. Li and X.N. Wang, nucl-th/0110075.

\bibitem{eks}
K.J. Eskola, V.J. Kolhinen, and C.A. Salgado, 
Eur. Phys. J. C {\bf 9}, 61 (1999).

\bibitem{HKM}
M. Hirai, S. Kumano, and M. Miyama, Phys. Rev. D {\bf 64}, 034003 (2001).

\bibitem{eskola02}
K.J. Eskola, H. Honkanen, V.J. Kolhinen, and C.A. Salgado, hep-ph/0201256
(2002). 

\bibitem{Davis}
C.T.~Davies, B.R.~Webber and W.J.~Stirling,
Nucl. Phys. {\bf B256}, 413 (1985).

\end{references}
\end{document}